\documentclass{ws-ijmpcs}
\usepackage{amssymb,graphicx}
\usepackage{epstopdf}
\DeclareGraphicsExtensions{.pdf,.eps,.png,.jpg,.mps}
\begin{document}
\markboth{L.T Handoko}
{Field theory approach in the dynamics of biomatter}

%
\catchline{}{}{}{}{}
%

\def\grad{\mathbf{\nabla}}
\def\om{\mathbf{\omega}}
\def\v{\mathbf{v}}
\def\x{\mathbf{x}}
\def\u{\mathbf{U}}
\def\f{\mathbf{F}}
\def\s{\mathbf{S}}
\def\en{\mathbf{E}}
\def\r{\mathbf{r}}
\def\mb{\bar{m}}
\def\mt{\tilde{m}}
\def\lt{\tilde{\lambda}}
\def\xp{{x^\prime}}
\def\l{{\cal L}}
\def\cd{{\cal D}}
\def\h{{\cal H}}
\def\f{{\cal F}}
\def\z{{\cal Z}}
\def\s{{\cal S}}
\def\e{{\cal E}}
\def\p{{\cal P}}
\def\cd{{\cal D}}
\def\cj{{\cal J}}
\def\pd{\partial}
\def\d{\mathrm{d}}
\def\j{\mathbf{J}}
\def\jt{\tilde{J}}
\def\lt{\tilde{\lambda}}
\def\mt{\tilde{m}}
\def\at{\tilde{\alpha}}
\def\jtv{\mathbf{\tilde{J}}}
\def\exp{\mathrm{exp}}
\def\ex{\mathrm{e}}
\def\be{\begin{equation}}
\def\ee{\end{equation}}
\def\bea{\begin{eqnarray}}
\def\eea{\end{eqnarray}}
\def\ie{\textit{i.e.} }
\def\etal{\textit{et.al.} }

\title{\bf FIELD THEORY APPROACH IN THE DYNAMICS OF BIOMATTER}

\author{L.T. HANDOKO}
\address{Group for Theoretical and Computational Physics,
Research Center for Physics, Indonesian Institute of Sciences,
Kompleks Puspiptek Serpong, Tangerang, Indonesia}
\address{Department of Physics, University of Indonesia,
Kampus UI Depok, Depok 16424, Indonesia\\
handoko@teori.fisika.lipi.go.id, handoko@fisika.ui.ac.id,
laksana.tri.handoko@lipi.go.id}

\maketitle

\begin{history}
\received{Day Month Year}
\revised{Day Month Year}
\end{history}

\begin{abstract}
A new approach to model the biomatter dynamics based on the field theory is presented. It is shown that 
some well known tools in field theory can be utilized to describe the physical phenomena in life matters, 
in particular at elementary biomatters like DNA and proteins. In this approach, the biomatter dynamics 
are represented as results of interactions among its elementary matters in the form of lagrangian. 
Starting from the lagrangian would provide stronger underlying theoretical consideration for further extension.  Moreover, it also enables us to acquire rich physical observables using statistical mechanics instead of 
relying on the space-time dynamics from certain equation of motions which is not solvable due to its 
nonlinearities. Few examples from  previous results are given and explained briefly.

\end{abstract}

\ccode{PACS numbers: 87.15.ad, 87.15.Cc, 87.15.hm}

\section{Introduction}

Presently, the human understanding of living things are still very far from perfect when compared to the 
elementary particle physics. In elementary particle physics all elementary constituents of matters and its 
interactions are relatively well known. They have uniquely and successfully been modeled in certain mathematical 
representations in a unified manner. Though the model, namely the standard model of particle physics, is still 
lacking from some major issues like the massive neutrinos, the origin of symmetry breaking leads to 
mass differences between particles. However those problems do not reduce its success story of modeling 
the huge numbers of phenomena in our nature in a single framework. This success is merely based on 
the principle of symmetry and a great tool called as field theory \cite{ryder}. The symmetry has mainly been introduced 
to restrict the model under the basic assumption that the theory should be invariant under time and spatial 
changes. On the other hand, field theory plays an important role across the concept, from the initial 
lagrangian till the calculation of real phenomenological observations \cite{feynman,rattazzi}.

Another notable reason of the success story in particle physics is the exploration of elementary matters. 
The particle physicists have for decades investigated the most fundamental building blocks of our nature. 
This effort ends up, at least for time being, at 16 elementary particles with the additional one hypothetic 
Higgs particles, currently investigated at the Large Hadron Collider (LHC) experiment at CERN \cite{pdg}. 
It finally can be concluded that, constructing a unified model is getting easier if one considers only 
few numbers of a unified framework is possible when one goes to the elementary scale with relatively 
small number of interacting matters. Even in the scale of nuclear matters at $\sim 10^{-15}$ m which involved 
hundreds of nuclear species, there is no reliable and unified model to describe its phenomena.

In contrast with the 'phenomenal progress' in particle physics, living creatures involve much more complex 
systems. Although all matters should be formed from the above 16 elementary particles, the living 
characteristics emerges due to the 'elementary' living building blocks with much larger sizes. If one 
treats the protein, DNA / RNA, etc as the elementary biomatters, then we are faced with the interactions 
of 'point particles' with the square-rooted size of quarks in particle physics. Note that the quark size is $\sim 10^{-18}$ m, while protein or DNA is at nanometer scale, \ie one order larger than an atom size.

Now let us consider the protein or DNA as the elementary biomatters. Although its number is still huge compared 
to the 16 elementary particles, they are the most fundamental building blocks of living things have been observed. 
The problem is then turned out to the applicability of particle physics methods. The main issue is the scale 
difference, in particular for using field theory. Because in the field theory one has a basic assumption that 
each interacting particle is treated as a point particle without any geometrical shapes. This is clearly not 
the case for biomatters. However, in physics one always considers the relative size rather than the absolute 
ones. If one can make use of field theory on the particle size matter to explain the observed nuclear phenomenology 
(all elementary particle observations are through the nuclear, \ie hadronic, processes), then it is also 
plausible to apply it on the biomatters to describe more macroscopic ensemble of living things. Alternatively, 
one can also bring an ad-hoc assumption to model the macroscopic ones using much more microscopic, but 
existing modelization. The second approach is commonly adopted among nonlinear theorists to deal with nonlinear 
phenomena like sea wave using nonlinear Schrodinger or even relativistic Klein-Gordon equations.

Assuming that the biomatter interactions can be modeled in certain lagrangians, one can further derive not 
only the governing equation of motion (EOM) describing the mechanical behavior, but also extract relevant 
physical observables along the time or temperature evolution 
through statistical mechanics methods \cite{bellac}. It should be remarked that the method is 
reliable only if the system is at an equilibrium state. 
Similar to the case of particle physics, it means 
the obtained physical observables are valid only for a particular period of equilibrium time.

In this paper, few examples of modeling the biomatters using field theory approach are presented. Beside 
discussing the derived EOMs and its behavior, some interesting results from statistical mechanics 
calculations are also given.

The paper is organized as follows. After this introduction, two examples of constructing lagrangians for 
the case of DNA and protein dynamics are presented. It is followed by the discussion on the statistical 
mechanics calculation to extract appropriate physical observables is given in Sec. 3. Finally the paper 
is ended with a short conclusion.

\section{The dynamics of biomatter}

The dynamics of biomatter has been worked out since long time ago. In most variant of models, the dynamics 
are described by certain EOM. For protein the pioneering works were conducted by Davydov \etal, while Yakushevich, 
Peyrard-Dauxois-Bishop \etal applied it into the case of DNA system \cite{davydov,yakushevich,peyrard}. 
Most of works have deployed the classical Hooke’s force of pendulum to construct the nonlinear EOM with a 
solitonic solution of oscillator harmonics.

As already mentioned in the preceding introduction, this approach obviously has disadvantages of 
lacking of first principle which subsequently leads to no clue for expanding the theory to incorporate 
more complicated dynamics.

There are some improved modelings yet based on the  oscillator harmonics system, but in the form of 
hamiltonian. The new interactions representing either internal or environmental effects are incorporated 
as potential terms. Using the hamiltonian, one can derive analytically the physical observables 
through its partition function using statistical mechanics. These methods predict interesting phenomena, 
for instance the anomaly of negative heat capacity in the case of monomer with anharmonic oscillation 
effects in thermal bath. The anomaly has been observed at certain region of lower 
finite physiological temperature using Linblad formulation \cite{sulaiman2}. 

Furthermore, the thermodynamic properties of Davydov-Scott monomer contacting with thermal bath has 
also been investigated using Lindblad open quantum system formalism through path integral method. It has
been found that the environmental effects contribute destructively to the specific heat, and large interaction between amide-I and amide-site is not preferred for a stable Davydov-Scott monomer \cite{sulaiman7}.

Unfortunately, although represented in non-relativistic hamiltonians, there is still no clue on how to 
expand further the hamiltonian. Because there is no basic assumption which is generally valid to restrict 
ourself when introducing the new terms of interaction. Therefore, we have proposed a new model based on 
the lagrangian with certain symmetry. The model was originally motivated by a need to take into account 
the effects from surrounding fluid, since the living systems are always immersed in a fluid medium. 
Such models are already available, one of them is the so-called fluid QCD representing the quark-gluon-plasma 
(QGP) system of highly densed and strongly interacting plasma \cite{handoko}. The model has been applied to 
describe the high energy QGP and plasma dominated early universe \cite{handoko2,handoko3}. The model is 
justified by the fact that it induces the non-relativistic Euler equation with certain external forces when 
one takes the non-relativistic limit \cite{handoko}.

Concerning the fact that an (elementary) biomatter has no intrinsic degree of freedom like
spin, it is considerable to represent its elementary constituents as  scalar
(boson) fields governed by the bosonic lagrangian, 
\be
        \l_\mathrm{matter} = \left( \pd_\mu \Phi \right)^\dagger  \left( \pd^\mu \Phi \right) + V(\Phi) \; , 
        \label{eq:lphi}
\ee
where $V(\Phi)$ is the potential. For example in the typical $\Phi^4-$theory, 
\be
	V(\Phi) = -\frac{1}{2} m_\Phi^2 \, \Phi^\dagger \Phi - \frac{1}{4!} \lambda \, (\Phi^\dagger \Phi)^2 \; ,
	\label{eq:v}
\ee
where $m_\Phi$ and $\lambda$ are the mass of matter and the dimensionless coupling constant of matter self-interaction. One can impose the above bosonic lagrangian to be gauge invariant under local (in general non-Abelian) gauge transformation, $U \equiv \exp[-i T^a \theta^a(x)] \approx 1 - i T^a \theta^a(x)$ with $\theta^a \ll 1$. $T^a$'s are generators belong to a particular Lie group and satisfy certain commutation  relation $[T^a,T^b] = i f^{abc} T^c$ with $f^{abc}$ is the anti-symmetric structure constant. The matter field is then transformed as $\Phi \stackrel{U}{\longrightarrow} \Phi^\prime \equiv \exp[-i T^a \theta^a(x)] \, \Phi$, with $T^a$ are $n \times n$ matrices while $\Phi$ is an $n \times 1$ multiplet containing $n$ elements, \ie 
\be
	\Phi = \left( 
	\begin{array}{c}
		\Phi_1 \\
		\Phi_2 \\
		\vdots \\
		\Phi_n \\
	\end{array}
	\right) 
	\; \; \; \mathrm{and} \; \; \; 
	\Phi^T = (\Phi_1 \; \Phi_2 \; \cdots \, \Phi_n) \; ,
	\label{eq:multiplet}
\ee
for $n$ dimension Lie groups as SU($n$), O($n+1$),  etc. It is well-known that the symmetry in Eq. (\ref{eq:lphi}) is revealed by introducing gauge fields $A_\mu^a$ which are transformed as $U^a_\mu \stackrel{U}{\longrightarrow} {U^a_\mu}^\prime \equiv U^a_\mu - \frac{1}{g}  (\pd_\mu \theta^a) + f^{abc} \theta^b U^c_\mu$, and replacing the derivative with the covariant one, $\cd_\mu \equiv \pd_\mu + i g \, T^a U^a_\mu$. Anyway, the number of generators, and also gauge bosons, is determined by the dimension of group under consideration. For an SU($n$) group one has $n^2 - 1$ generators and the index $a$ runs over $1, 2, \cdots, n^2 - 1$.

The gauge boson $U_\mu$ is  interpreted as a ``fluid field'' with velocity
$u_\mu$, and takes the form $U^a_\mu = \left( U_0^a, \u^a \right) \equiv u^a_\mu \, \phi$ with 
$u_\mu \equiv \gamma^a (1, -\v^a)$ and $\phi$ is an auxiliary boson field, while 
$\gamma^a \equiv \left( 1 - |\v^a|^2 \right)^{-1/2}$ \cite{handoko,handoko2}.
This is nothing else than rewriting a gauge field in terms of
its polarization vector and wave function which represents the fluid
distribution in a system. It has further been shown that the non-relativistic
fluid equation can be reproduced \cite{handoko2}.

For the simplest case of single matter to represent the single strand of DNA, 
using the Euler-Lagrange equation one obtains an EOM as follow \cite{sulaiman}, 
\be
	\left( \partial^2 + m_\Phi^2 + 2 g^2 \, U^2 \right) \Phi + \frac{1}{3!} \lambda \, \Phi^3 = 0 \; .
	\label{eq:eomu1}
\ee
for a real $\Phi$ field. This can be solved analytically and the solution is well known. It can be concluded 
that the surrounding fluid enhances the amplitude at short distance \cite{sulaiman}. Moreover, the double 
stranded DNA can be modeled using the same lagrangian but with $n=2$. Again, the Euler-Lagrange equation 
 in this case yields the following EOM \cite{sulaiman}, 
\be
	\left( \partial^2 + m_\Phi^2 -4 i g \, \sigma_2 U_2^\mu \partial_\mu \right) \Phi + \frac{1}{3!}\lambda \, \left( \Phi^T \Phi \Phi \right) = 0 \; ,
	\label{eq:eomsu2}
\ee
at non-relativistic limit for constant fluid velocity and $\phi$. Here, $\Phi^T = (\phi_1,\phi_2)$. This can 
be solved analytically under further assumption as done in  \cite{sulaiman}.

Anyway, so far we have concerned only the EOM's from the lagrangian. All of these would end up to the nonlinear 
differential equations which are in most cases unsolvable. Therefore one should consider another way to extract 
meaningful physical observables.

\section{Statistical mechanics observables}

For the sake of simplicity let us consider a simple example of protein dynamics using the 
so-called $\phi^4$ theory \cite{sulaiman6}. The calculation has been performed using statistical mechanics 
and path integral method. In particular, the evolution of heat capacity in term of temperature has been 
investigated for various levels of the nonlinearity of source, and the strength of interaction between protein backbone and nonlinear source. It has been argued that the nonlinear source contributes constructively to the specific heat especially at higher temperature when it is weakly interacting with the protein backbone. 
This may indicate increasing energy absorption as the intensity of nonlinear sources are getting greater.

Similar to the previous case, the model adopts the $\phi^4$ lagrangian but with additional interactions \cite{sulaiman6},
\be
\l_{tot} = \l_c(\phi)+\l_s(\psi)+\l_{int}(\phi,\psi)\; ,
\label{eq:ltot}
\ee
where, 
\bea
\l_c	& = & \frac{1}{2}\left[ \left( \pd_\mu \phi \right)^\dagger \left(
\pd^\mu \phi \right) + m_{\phi}^2 \left( \phi^\dagger \phi \right) \right] \; ,
\label{eq:Lc}\\
\l_s	& = & \frac{1}{2} \left[ \left( \pd_\mu \psi \right)^\dagger  \left(
\pd^\mu \psi \right) - \frac{1}{2} \lambda \, \left( \psi^\dagger \psi
\right)^2 \right] \; , \label{eq:Ls}\\
\l_{int} & = & \Lambda \, \left( \phi^\dagger \phi \right) \left( \psi^\dagger
\psi \right) \; ,
\label{eq:Li}
\eea
representing the conformational changes of a protein backbone and the nonlinear
source injected to the backbone, while the last one is the interaction term 
between both. Imposing a local U(1) symmetry to the total lagrangian and considering its
minima lead to the vacuum expectation value (VEV), 
$\langle \psi \rangle = \sqrt{{2\Lambda}/{\lambda}} \langle \phi \rangle$. 
This non-zero VEV then yields the so-called spontaneous symmetry breaking.
The lagrangian yields two EOM's in term of $\phi$ and $\psi$ respectively. However we are 
not interested in discussing its dynamical behaviors. The statistical observables can be 
conveniently calculated from the generating functional by the perturbation method 
\cite{ryder}. The generating functional for scalar fields is written as,
\be
Z = \int \mathcal{D}\phi\mathcal{D}\psi \, \exp\left\{ 
\int_0^{\beta} \int_0^L d\tau dx \, \l_{tot}(\phi,\psi)\right\}\;,
\label{eq:generate}
\ee
at the finite temperature case in Euclidean coordinates.
The integral in Eq. (\ref{eq:generate}) can be evaluated analytically using the
Gaussian integral. This can be accomplished by rewriting it in term of Gaussian
integral using the Fourier representation of Green's function to obtain the 
master equation \cite{sulaiman6},
\bea
Z & = & N \exp \left\{ \int_0^{\beta} d\tau \int_0^L dx 
\left[ -\frac{3}{4}\lambda\Delta_{\psi}^2(0) +  
\Lambda\Delta_{\phi}(0)\Delta_{\psi}(0)\right]\right\} \; .
\eea

For statistical observables, Let us consider the specific heat of the system in a constant volume, 
$C_V$, that is a particular interest from experimental point of view. The specific heat
can be derived directly from the partition function using the relation, 
\be
C_V = \beta^2 \, \left( \frac{\partial^2 \ln Z}{\partial \beta^2} \right)_V \; . 
\ee
In the present case, it has been found to be \cite{sulaiman6},
\bea
C_V & = & \beta^2 \, \frac{\partial^2}{\partial \beta^2} \left(\ln{N} - 
 \beta L \left[ \frac{3}{4}\lambda\Delta_{\psi}^2(0)
- \Lambda\Delta_{\phi}(0)\Delta_{\psi}(0)\right]\right) \; ,
\label{eq:omega}
\eea
after performing the integration over $\tau$ and $x$ respectively, while the
overall factor $N = {[ 4 \pi \sinh({k \beta}/2) ]}^{-1}$.

The integration in Eq. (\ref{eq:omega}) should be done numerically. It has been argued that 
both $\lambda$ and $\Lambda$ contribute to the evolution of $C_V$ in an
opposite way, and could completely cancel each other at certain values. 
This occurs when the symmetry is maximally broken which means that increasing energy absorption 
prefers high level of nonlinearity of sources and at the same time weak interaction between
the sources and protein backbone.

\section{Conclusion}

New models for biomatter dynamics using lagrangian have been presented. The approach have been 
applied to describe typical cases of  single and double stranded DNA, and the protein folding mechanism. 
The usage of statistical mechanics method to extract the specific heat has also briefly been 
demonstrated. It can be argued that the approach might be able to provide new directions on 
modeling the living phenomena at elementary level.  

However more theoretical works are still going on. More importantly it should be justified by 
experiments in the future. 

\section*{Acknowledgments}

LTH greatly appreciate the Humboldt Foundation for financial support, and the Organizer 
for warm hospitality during the conference. 
LTH also acknowledges long collaborators in the topics, in particular A. Sulaiman and M. Januar for their 
passion and hard works. This work is funded by the Indonesia Ministry
of Research and Technology and the Riset Kompetitif LIPI in fiscal year 2011
under Contract no.  11.04/SK/KPPI/II/2011.

\bibliographystyle{ws-procs975x65}
\bibliography{jagna}

\begin{thebibliography}{10}

\bibitem{ryder}
L.~H. Ryder, {\em Quantum Field Theory 2nd Ed.} (Cambridge University Press,
  1996).

\bibitem{feynman}
R.~P. Feynman and A.~R. Hibbs, {\em Quantum Mechanics and Path Integrals}
  (McGraw-Hill, 1965).

\bibitem{rattazzi}
R.~Rattazzi, {\em The Path Integral Approach to Quantum Mechanics} (EPFL,
  2009).

\bibitem{pdg}
K.~N. et~al. (Particle Data~Group), {\em Journal of Physics G} {\bf 37}, p.
  075021 (2010).

\bibitem{bellac}
M.~L. Bellac, {\em Thermal Field Theory} (Cambridge University Press, 1996).

\bibitem{davydov}
A.~S. Davydov, {\em Solitons in Molecular Systems} (Kluwer, 1981).

\bibitem{yakushevich}
L.~V. Yakushevich, {\em Nonlinear Physics of DNA} (Wiley \& Sons, 1998).

\bibitem{peyrard}
M.~Peyrard and A.~R. Bishop, {\em Nonlinearity} {\bf 17}, R1 (2004).

\bibitem{sulaiman2}
A.~Sulaiman, F.~P. Zen, H.~Alatas and L.~T. Handoko, {\em Physical Review E}
  {\bf 81}, p. 061907 (2010).

\bibitem{sulaiman7}
A.~Sulaiman, F.~P. Zen, H.~Alatas and L.~T. Handoko, The thermodynamic
  properties of davydov-scott's protein model in thermal bath, in {\em
  Proceeding of the Conference in Honour of Murray Gell-Mann's 80th Birthday :
  Quantum Mechanics, Elementary Particles, Quantum Cosmology, Complexity\/},
  2011.

\bibitem{handoko}
A.~Sulaiman, A.~Fajarudin, T.~P. Djun and L.~T. Handoko, {\em International
  Journal of Modern Physics A} {\bf 24}, 3630 (2009).

\bibitem{handoko2}
T.~P. Djun and L.~Handoko, Fluid {QCD} approach for quark-gluon plasma in
  stellar structure, in {\em Proceeding of the Conference in Honour of Murray
  Gell-Mann's 80th Birthday : Quantum Mechanics, Elementary Particles, Quantum
  Cosmology, Complexity\/},  {\em Proceeding of the Conference in Honour of
  Murray Gell-Mann's 80th Birthday : Quantum Mechanics, Elementary Particles,
  Quantum Cosmology, Complexity} 2011.

\bibitem{handoko3}
C.~S. Nugroho, A.~O. Latief, T.~P. Djun and L.~T. Handoko, {\em Gravitation and
  Cosmology} {\bf 18}, 32 (2012).

\bibitem{sulaiman}
A.~Sulaiman and L.~T. Handoko, {\em Journal of Computational and Theoretical
  Nanoscience} {\bf 8}, 124 (2011).

\bibitem{sulaiman6}
M.~Januar, A.~Sulaiman and L.~T. Handoko, Conformation changes, protein folding
  induced by $\phi^4$ interaction, in {\em Proceeding of the Conference in
  Honour of Murray Gell-Mann's 80th Birthday : Quantum Mechanics, Elementary
  Particles, Quantum Cosmology, Complexity\/}, 2011.

\end{thebibliography}

\end{document}